\pdfoutput=1
\documentclass[aip,preprint,amsmath,amssymb,aps,showkeys]{revtex4-2}

\usepackage{graphicx}
\usepackage{dcolumn}
\usepackage{bm}
\usepackage{xcolor}

\usepackage{soul}										
\setstcolor{red}    
\usepackage[normalem]{ulem}  							    

\newcommand{\nuc}[2]{\ifmmode{{}^{#2}\mathrm{#1}}\else{${}^{#2}\mathrm{#1}$}\fi}

\begin{document}

\preprint{APS/123-QED}

\title{Density filamentation nonlinearly driven by the  Weibel instability in relativistic beam-plasmas}

\author{Cong Tuan Huynh}
\affiliation{Center for Relativistic Laser Science, Institute for Basic Science (IBS), Gwangju 61005, Republic of Korea}

\author{Chang-Mo Ryu}%
 \email{Corresponding author: ryu201@postech.ac.kr}
\affiliation{Center for Relativistic Laser Science, Institute for Basic Science (IBS), Gwangju 61005, Republic of Korea}
\affiliation{Department of Physics, POSTECH, Pohang 37673, Republic of Korea}

\author{Chulmin Kim}
\affiliation{Center for Relativistic Laser Science, Institute for Basic Science (IBS), Gwangju 61005, Republic of Korea}
\affiliation{Advanced Photonics Research Institute, Gwangju Institute of Science and Technology, Gwangju 61005, Republic of Korea}

\date{\today}

\begin{abstract}
Density filamentation has  been observed in many beam plasma simulations and experiments. Because  current  filamentation is a  pure transverse mode, charge density filamentation cannot be produced directly by the current filamentation process.  To explain this phenomenon, several mechanisms are proposed such as the coupling of the Weibel instability to the two-stream instability, coupling  to the Langmuir wave,    differences in thermal velocities between the beam and return currents, the magnetic pressure gradient force, etc.   In this paper, it is shown that the gradient of the Lorentz factor can, in fact, represent the nonlinear  behavior of a plasma fluid, and further that the nonuniform  Lorentz factor distribution can give rise to electrostatic fields and  density filaments.  Simulation results together with theoretical analyses are presented.
\end{abstract}

\maketitle



\section{Introduction}
\label{s01}
Collisionless shocks are important in many astrophysical bodies such as  gamma-ray bursts, supernova remnants, active galactic nuclei, and pulsar wind nebulae.  Synchrotron radiation observed from these astrophysical bodies typically characterizes the existence of strong magnetic fields and highly accelerated particles far beyond  thermal energy. Numerical simulations have demonstrated that the Weibel/filamentation instability can play an important role in forming the collisionless shock.  \citep{Nishikawa2003, Frederiksen2004, Spitkovsky2008}  The Weibel/filamentation instability, which is generally called as filamentation instability (FI), is now  well understood. \citep{Weibel1959, Fried1959, Medvedev1999, Schlickeiser2005, Fiuza2012, Fox2013, Huntington2015} However, it still remains unclear how upstream electrons get energy in the foreshock region. \citep{Kumar2015} Numerical simulations have shown that there exists a large-scale longitudinal electric field occurring around current filaments. The Weibel/filamention instability inducing current filaments is a transversal mode and cannot effectively heat  upstream electrons. In fact, the analysis of  numerical PIC (Particle-In-Cell) simulation results has demonstrated that most of the heating comes from a small amount of longitudinal electric fields than more prevalent transversal fields. \citep{Kumar2015}

However, in many experimental and simulation studies of relativistic beam-plasmas,  electron density filamentation has been observed with accompanying electrostatic fields.  \citep{Honda2000,Nishikawa2003,Tatarakis2003,Frederiksen2004,Grassi2017,Kumar2015}  In the simulation  study of inertial fusion problems, using high-intensity laser,  density filamentation has been observed during the formation of current filaments. \citep{Honda2000} In 3D collisionless shock simulation, it was  noticed that  the electron density filamentation occurs at the jet front, and  that  accompanying strong electrostatic fields  accelerate electrons and ions. \citep{Nishikawa2006, Nishikawa2009}

The Weibel/filamentation instability is a pure transverse mode and thus cannot simply produce charge-density filaments. \citep{Bret2005} To explain the   phenomenon of density filamentation and electrostatic fields observed in many beam-plasma  simulations with dominant transverse modes, several mechanisms are proposed.     Pegoraro et al. suggested that  the Weibel instability can  couple to the Langmuir wave in the nonlinear regime. \citep{Pegoraro1996}   Bret et al. argued that  because the current carried by the beam is susceptible to the two-stream instability, and that the  two-stream instability can combine with the Weibel/filamentation mode,   a mixed mode of longitudinal and transverse modes could  be excited. \citep{Bret2005, Bret2016}  It was found that the growth rate  mode of  a mixed mode propagating obliquely to the beam direction with a quasi-longitudinal electric field was greater than that of the pure transverse Weibel/filamentation mode. 
 As a variation to this approach, Tzoufras et al.  suggested that forward and backward flowing electrons can have different transverse temperatures, resulting the total distribution function  not to be separable, and thus that different magnetic pinching between the beam and the return currents can cause  density filamentation. \citep{Tzoufras2006}  In their study, it is found that the growth rate of the filamentation instability is substantially reduced due to the coupling to  density filamentation.   In another alternative approach, Rowlands et al. have  shown that using a one dimensional study of nonlinear evolution of the filamentation instability driven by cool nonrelativsitic beams, although  electrostatic fields cannot be generated  by the FI in the linear phase,   the magnetic pressure gradient force  can be built up with the growth of the instability, which  can in turn induce a large amount of electrostatic fields. \citep{Rowlands2007, Dickmann2009, DickmannBret2009} They have shown that the gradient of the magnetic pressure, which develops during the quasi-linear evolution of the filamentation instability, can give rise to an electrostatic field.

In this  paper,  we investigate the generation of the electrostatic field by the nonlinear filamentation instability for a relativistic beam plasma, expanding Rowlands et al.'s demonstration of the electrostatic field generation by quasi-linear effects in a nonrelativistic plasma.    We first show that the  nonlinear plasma behavior  can be expressed in terms of the gradient of the Lorentz factor, relating the nonlinear fluid motion to the gradient of the Lorentz factor, and the latter can  be related  to the density filamentation.   We  present an analytical theory and  supporting  simulation results obtained using the  PIC (Particle-In-Cell) code EPOCH. \citep{Arber2015}

\section{Electron Density Filamentation by Nonlinear  Fluid Motion}
\label{s02}
\begin{figure}[h]
		\centering
			\includegraphics[width=8 cm]{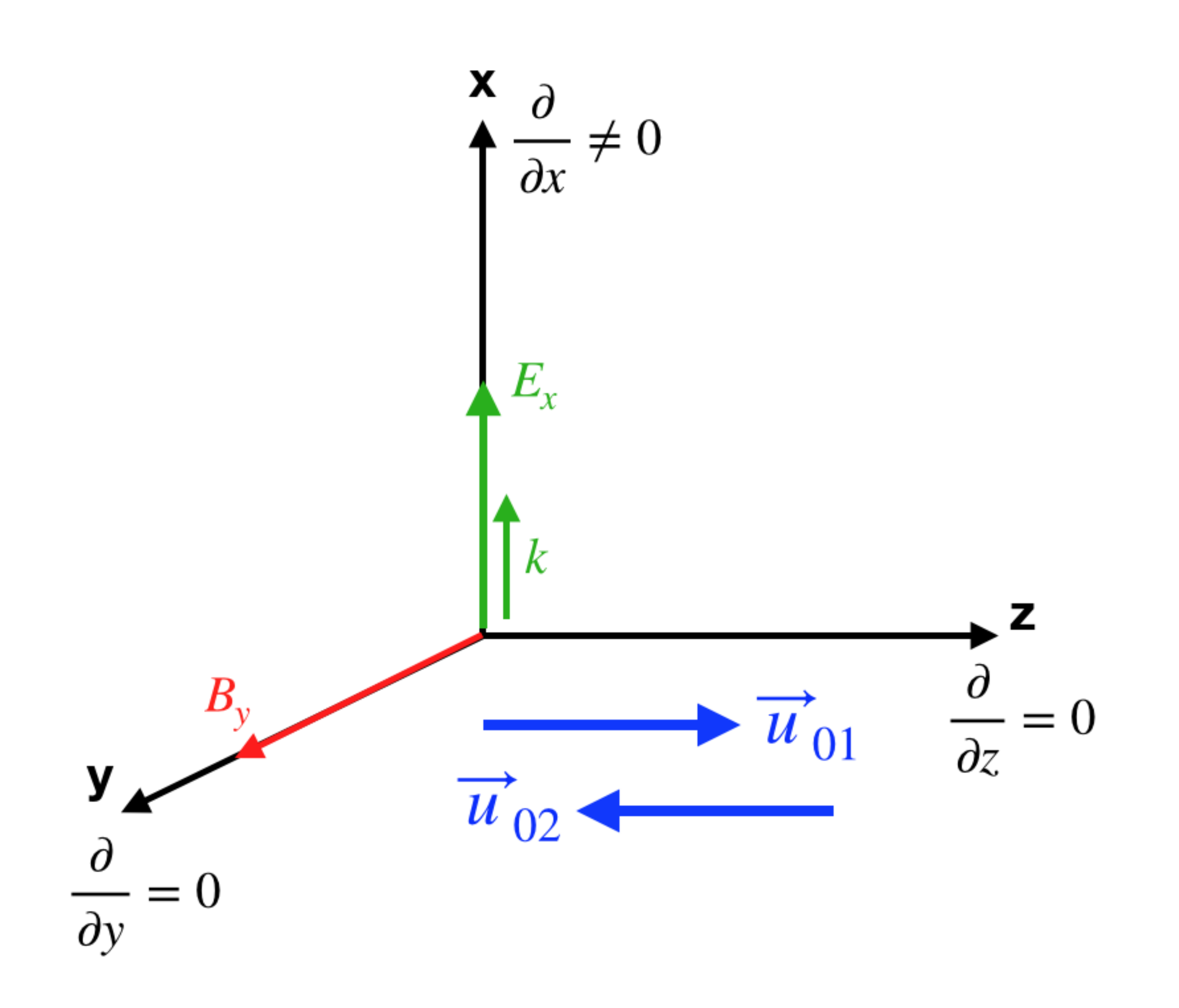}
		\caption{Schematic diagram of the geometry used in the study of 1D density filamentation  }
		\label{}
	\end{figure}

	For convenience, we shall normalize all physical quantities as follows: $\widetilde{t}=t\omega_{pe}$, $\widetilde{x}=\omega_{pe}x/c$, $\widetilde{n}=n/n_0$, $\widetilde{\textbf{u}}= \textbf{u}/c$, $\widetilde{\textbf{p}}=\textbf{p}/mc$, $\widetilde{\textbf{B}}=e\textbf{B}/m\omega_{pe}$, $\widetilde{\textbf{E}}=e\textbf{E}/mc\omega_{pe}$, and $\widetilde{\textbf{A}}=e\textbf{A}/mc$, where c is the speed of light, $\omega_{pe}=\sqrt{n_0e^2/\epsilon_0m_e}$ (in the SI units) is the total electron plasma frequency, $n_0$ is the initial number density of  the  total electron, $e$ is the elementary charge, $\epsilon_0$ is the vacuum permittivity, $m_e$ is the mass of the electron and $\textbf{A}$ is the vector potential. The tildes are dropped in the following. For simplicity, we ignore the pressure terms. The fluid-dynamic equation of the electron is given by
	\begin{equation}
		\frac{\partial \textbf{p}_{\alpha}}{\partial t} + \left(\textbf{u}_{\alpha} \cdot \nabla \right) \textbf{p}_{\alpha} = -\textbf{E} - \textbf{u}_{\alpha}\times\textbf{B}, \label{4eq1}
	\end{equation}
	where $\alpha$ represents the electron flow, $\textbf{p}_{\alpha}=\gamma_{\alpha} \textbf{u}_{\alpha}$ is the momentum and $\gamma_{\alpha}=1/\sqrt{1-\textbf{u}_{\alpha}^2}$ is the Lorentz factor. Applying the operator $\nabla \times$ on both sides of Eq.\,(\ref{4eq1}) and Faraday's law $\nabla\times \textbf{E}=-\partial \textbf{B} / \partial t$, we obtain
    \begin{equation}
		\frac{\partial \left(\nabla \times\textbf{p}_{\alpha}\right)}{\partial t} - \nabla \times \left[\textbf{u}_{\alpha} \times \left(\nabla \times \textbf{p}_{\alpha}\right) \right]  = \frac{\partial \textbf{B}}{\partial t} - \nabla \times \left[\textbf{u}_{\alpha}\times\textbf{B}\right]. \label{4eq2}
	\end{equation}

We assume  that there is  initially no vortical motion so that London's equation  $\textbf{B}=\nabla \times \textbf{p}_{\alpha}$ can be  obtained  from Eq.\,(\ref{4eq2}) (see Ref. \onlinecite{Kaganovich2001}). By inserting this relation into Eq.\,(\ref{4eq1}), the equation of motion can be reduced to (see the Appendix),
    \begin{equation}
		\frac{\partial \textbf{p}_{\alpha}}{\partial t} = -\nabla \gamma_{\alpha}  -\textbf{E}. \label{4eq3}
    \end{equation}
One can see that
  $-\nabla\gamma_{\alpha}$  balances  the  electric field on the right-hand side of this equation, implying that variation of the Lorentz factor can cause the electric field to occur.
To see its effects on  the FI,  we consider  two electron beams  counter streaming  in the $z$-direction. The first electron beam denoted by $e_1^-$ flows with the momentum $p_0 >0$, and the second one  by $e_2^-$ with the  momentum $p_0 <0$. All physical quantities depend only on  $x$ ($\partial/\partial x \neq 0, \, \partial/\partial y =0, \, \partial/\partial z =0$).

Rewriting each term of Eq.\,(\ref{4eq3})  as $\partial p_{x1}/\partial t = -\partial \gamma_{1} / \partial x - E_x$, and $\partial p_{x2}/\partial t = -\partial \gamma_2 /\partial x - E_x$, and then taking  the sum  of these two equations, we  obtain the following electric field,
 \begin{equation}
		E_x = - \frac{\partial (p_{x1}+p_{x2})}{2 \partial t} - \frac{\partial  (\gamma_{1}+\gamma_{2})}{2 \partial x}. \label{4eq04}
	\end{equation}

For perfectly counter streaming beams (Fig. 1), the first term on the RHS disappears and only the Lorentz factor term remains.
What Eq.\,(\ref{4eq04}) indicates is that when the Weibel instability occurs for perfectly counter streaming beams, an electrostatic field can be generated by the nonlinear interaction of that instability with the  background plasma.  This interaction can be very small and negligible when the amplitude of the Weibel instability is small, but as the instability develops, it can play an important role.

In a normalized form, the total electron density can be written as  $n=1+ \delta n$, where $\delta n$ is the perturbed density. Putting (\ref{4eq04}) into the Poisson equation, the  total electron density in 1D can be written   as
	\begin{equation}
		n = 1 + \delta n \simeq 1+\Gamma_{max}\frac{\partial (p_{x1}+p_{x2})}{2 \partial x} + \frac{\partial^2 (\gamma_{1}+\gamma_{2})}{2 \partial x^2}, \label{4eq4}
	\end{equation}
where we have used an assumption $\partial /\partial t \simeq \Gamma_{max}$. %

 The maximum growth rate  $\Gamma_{max}$ applied to Eq.\,(\ref{4eq4}) is given by the FI obtained by linearizing the set of the density continuity equation,  the momentum conservation equation, and the Maxwell equations. \citep{Bret2010,Bret2016,Bret2017}
  Once the dispersion relation is obtained, the maximum growth rate  can be estimated from it. In our study,  plasmas are initially unmagnetized. We assume that all perturbed quantities are of the form $\delta f \exp (i\textbf{k}\cdot\textbf{r}-i\omega t)$, where $\textbf{k}\equiv (k_x,0,0)$ is normalized by $\omega_{pe}/c$ and $\omega$ is normalized by $\omega_{pe}$. For ${\bf p}= \gamma {\bf u}$ with ${\bf p} = {\bf p}_0 +\delta {\bf p}$, the perturbed momenta are given by
\begin{eqnarray}
 \delta {\bf p}_\alpha = \gamma_{0\alpha} \delta {\bf u}_{\alpha} + \delta \gamma_{\alpha} {\bf u}_{0\alpha} +  \delta \gamma_{\alpha} {\delta \bf u }_{\alpha}.
 \label{4eq5}
\end{eqnarray} 
Keeping the first order terms, the momentum perturbation can be written  in accordance with the ones obtained with the  conventional approach \citep{Godfrey1975, Bret2010}  as,
\begin{eqnarray}
\delta {\bf p}_{\alpha } = \gamma_{0\alpha} \delta {\bf u}_\alpha +  \left[ \frac{ {\bf u_{0\alpha}} \cdot \delta{\bf u}_\alpha}{(1-u_\alpha^2)^{3/2}} \right] {\bf u}_{0\alpha} =  \gamma_{0\alpha} \delta {\bf u}_\alpha + \gamma_{0\alpha}^3 ({\bf u}_{0\alpha} \cdot \delta {\bf u}_\alpha) {\bf u}_{0\alpha}
\label{4eq5_add}
\end{eqnarray}

 Then, from the momentum Eq.\,(\ref{4eq1}),

\begin{equation}\label{5eq0}
  \delta \textbf{p}_{\alpha} = i\frac{ \textbf{E}+\frac{\textbf{u}_{0\alpha}\times[\textbf{k}\times \textbf{E}]}{\omega}}{\textbf{k}\cdot\textbf{u}_{0\alpha}-\omega },
\end{equation}
where $\textbf{u}_{0\alpha}\equiv (0,0,u_{0\alpha})$.

From Eq.\,(\ref{4eq5_add}),
\begin{equation}\label{5eq1}
  \delta u_{x\alpha} = \frac{\delta p_{x\alpha}}{\gamma_{0\alpha}}, \; \; \delta u_{y\alpha} = \frac{\delta p_{y\alpha}}{\gamma_{0\alpha}}, \; \; \delta u_{z\alpha} = \frac{\delta p_{z\alpha}}{\gamma^3_{0\alpha}}.
\end{equation}

Each term of Eq.\,(\ref{5eq1}) can be expressed as, by using Eq.\,(\ref{5eq0}),
\begin{eqnarray}\label{5eq2}
  &\delta u_{x\alpha}& = -i\frac{ 1}{\omega \gamma_{0\alpha}} E_x  - i \frac{k_x u_{0\alpha} }{\omega^2\gamma_{0\alpha}} E_z, \nonumber \\ 
  &\delta u_{y\alpha}& = -i \frac{1} { \omega \gamma_{0\alpha}} E_y, \nonumber \\ 
  &\delta u_{z\alpha}& = -i \frac{1} { \omega \gamma^3_{0\alpha}} E_z.
\end{eqnarray}

Using the density continuity equation, the perturbed density can be written as
\begin{equation}\label{5eq3}
  \delta n_{\alpha} = - \frac{\textbf{k}\cdot \delta \textbf{u}_{\alpha}}{\textbf{k}\cdot \textbf{u}_{0\alpha}-\omega}n_{0\alpha} =
  -i \frac{k_x n_{0\alpha}}{\omega^2\gamma_{0\alpha}}E_x  - i \frac{k_x^2u_{0\alpha} n_{0\alpha}}{\omega^3 \gamma_{0\alpha}}  E_z.
\end{equation}

In the linear form, the current density is given by
\begin{equation}\label{5eq4}
  \delta \textbf{J} = -\sum_{\alpha}\left(n_{0\alpha}\delta \textbf{u}_{\alpha} + \delta n_{\alpha} \textbf{u}_{0\alpha}\right).
\end{equation}

Then, the  Maxwell equations  give
\begin{equation}\label{5eq5}
  \textbf{k}\times(\textbf{k}\times \textbf{E}) + \omega^2(\textbf{E}+i\delta \textbf{J}/\omega)=0.
\end{equation}

Inserting  the expressions obtained in Eqs.\,(\ref{5eq2}) and (\ref{5eq3}) in terms of $\textbf{E}$ into Eq.\,(\ref{5eq4}) to get $\delta \textbf{J}$, and then, using the final expression in Eq.\, (\ref{5eq5}), we obtain
\begin{equation}\label{5eq6}
  \left(
            \begin{array}{ccc}
              \epsilon_{xx} & 0 & \epsilon_{xz} \\
              0 & \epsilon_{yy} & 0 \\
              \epsilon_{zx} & 0 & \epsilon_{zz} \\
            \end{array}
  \right)
  \left(
    \begin{array}{c}
      E_x \\
      E_y \\
      E_z \\
    \end{array}
  \right) = 0,
\end{equation}
where
\begin{eqnarray}
  \epsilon_{xx} &\equiv& \omega^2-\sum_{\alpha}n_{0\alpha}/\gamma_{0\alpha} \nonumber \\
  \epsilon_{xz} &=& \epsilon_{zx} \equiv - k_x\sum_{\alpha}u_{0\alpha}n_{0\alpha}/(\omega\gamma_{0\alpha}) \nonumber \\
  \epsilon_{yy} &\equiv& \omega^2 - k_x^2 -\sum_{\alpha}n_{0\alpha}/\gamma_{0\alpha} \nonumber \\
  \epsilon_{zz} &\equiv& \omega^2 - k_x^2 -  k_x^2 \sum_{\alpha}u^2_{0\alpha} n_{0\alpha}/(\omega^2\gamma_{0\alpha}) - \sum_{\alpha}n_{0\alpha}/\gamma^3_{0\alpha} . \label{5eq7}
\end{eqnarray}

In the symmetric counter streaming case, $u_{01}=-u_{02}=u_0$, $\gamma_{01}=\gamma_{02}=\gamma_0$, and $n_{01}=n_{02}=1/2$, so $\epsilon_{xz}=0$ and the matrix in Eq.\,(\ref{5eq6}) becomes diagonal. Three dispersion relations, corresponding to three modes, are obtained; the Langmuir mode $\epsilon_{xx}=0$, the pure electromagnetic mode $\epsilon_{yy}=0$, and the filamentation mode $\epsilon_{zz}=0$. In Fig.2, the growth rate  with respect to the wave number is plotted for the velocities $u_0=0.0025, 0.6$ and $0.9$. Saturation of the growth rate is shown for both the nonrelativistic and relativistic cases. In the case of the relativistic beam, the growth rate is much higher than that in the nonrelativistic case. Taking into account of large $k$ values \citep{Grassi2017}, the maximum growth rate of the filamentation mode can be obtained as
\begin{equation}\label{5eq8}
  \Gamma_{max} = \sqrt{\sum_{\alpha} u^2_{0\alpha}n_{0\alpha}/\gamma_{0\alpha}} = u_0\sqrt{1/\gamma_{0}}.
\end{equation}

\begin{figure}[h]
	\centering
	\includegraphics[width=8 cm]{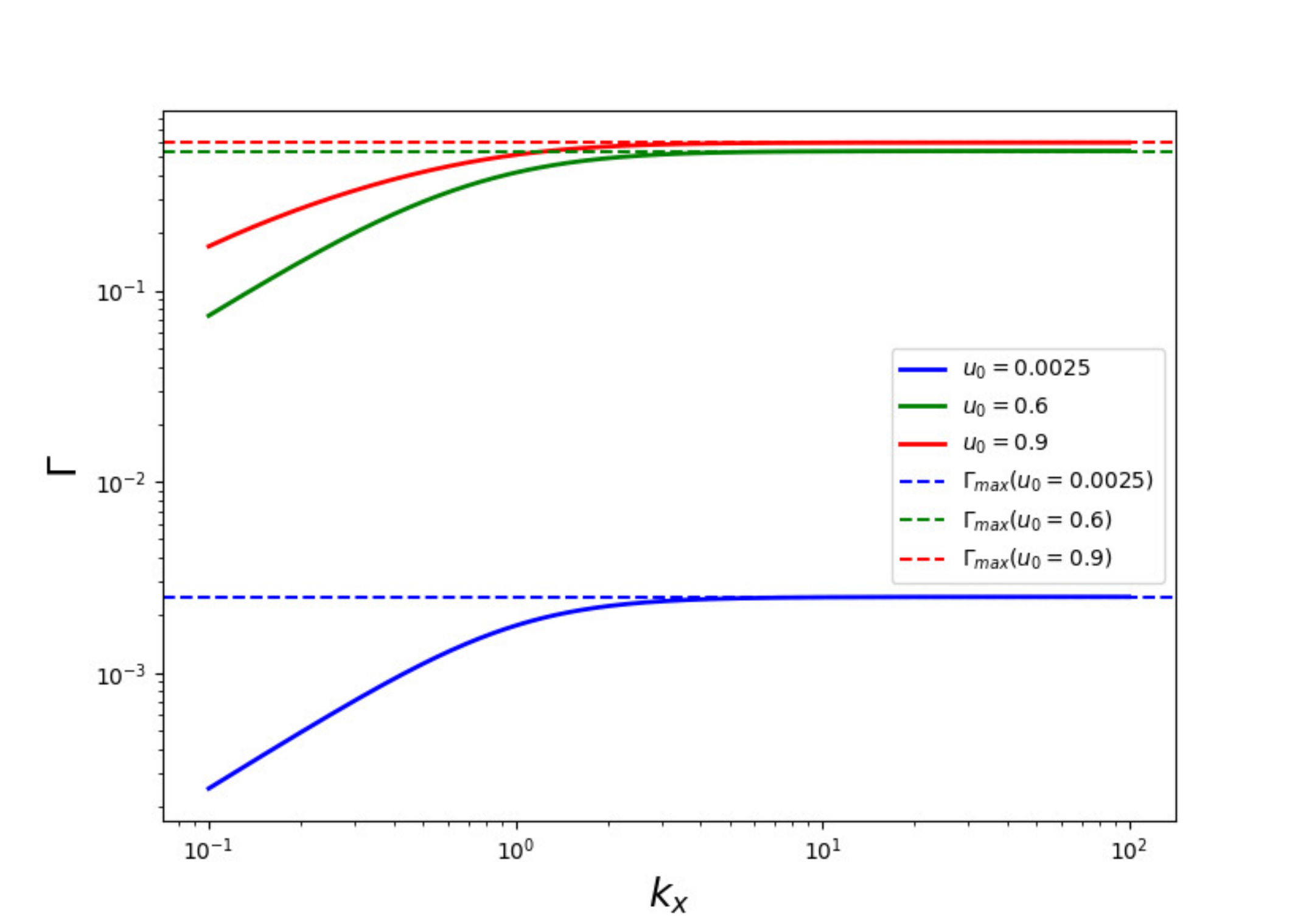}
	\caption{Filamentation growth rates  with respect to the wave numbers for different beam velocities.  Solutions are obtained from   $\epsilon_{zz}=0$ in Eq.\,\ref{5eq7}. }
	\label{}
\end{figure}

This growth rate is used in Eq.\,(\ref{4eq4}) to estimate the total electron density. By  linearization, it can be seen that the total electron density  perturbation by the symmetric counter streaming beams   comes from the two uncoupled modes; the Langmuir and  filamentation modes. The density perturbation by the Langmuir mode is not significant. The total electron density perturbation  triggered by filamentation instability is given by
\begin{equation}\label{5eq9}
  \delta n = \delta n_1 + \delta n_2 = -i \frac{k_x^2 }{\omega^3 }  E_z \left( \frac{n_{01}u_{01}}{\gamma_{01}} + \frac{n_{02}u_{02}}{\gamma_{02}}\right) = 0.
\end{equation}

This means that in the linear stage the symmetric counter streaming FI cannot trigger density filamentation even in the relativistic regime.

For small perturbations,   $\nabla \delta \gamma_\alpha$ in Eq.\,(\ref{4eq3}) can be  written as, by (\ref{eq_A11}),  
\begin{eqnarray}\label{5eq10}
\nabla \delta \gamma_\alpha = {\bf u}_{0\alpha} \times \delta {\bf B} +  ({\bf u}_{0\alpha} \cdot \nabla) \delta {\bf p}_\alpha  + {\delta \bf u}_{\alpha} \times \delta {\bf B}  + ({\delta \bf u}_{\alpha} \cdot \nabla) \delta {\bf p}_\alpha,
\end{eqnarray}
and thus, it can be seen that only the first two terms are used in the linear eigenmode analysis. The remaining two nonlinear terms can be neglected for small amplitude perturbations, but they can become important when the perturbation amplitudes increase.

In the initial phase of the nonrelativistic limit,   where the instability is not fully developed so that the amplitude of the wave is small, $\gamma_{1} \approx 1, \gamma_{2}\approx 1$, and thus $\partial^2 (\gamma_{1}+\gamma_{2})/\partial x^2 \approx 0$.  Because the FI has transverse electric fields, when the $e_1^-$ electrons are bunched during the development of the FI, the same amount of  $e_2^-$ electrons are expelled by Eq.\,(\ref{5eq9}).  This implies that $\partial (p_{x1}+p_{x2})/\partial x =0$, and that two electron beams  develop filamentation  at the same rate. Therefore, Eq.\,(\ref{4eq4}) gives $n=1$,  meaning that the total electron density remains uniform.  
  On the other hand, in the relativistic regime, the term $\partial^2 (\gamma_{1}+\gamma_{2})/\partial x^2$ may not vanish because of the third and  fourth  nonlinear terms in Eq.\,(\ref{5eq10}).  The nonlinear terms can cause filamentation  of the electron density, according to Eq.\,(\ref{4eq4}). 
  Therefore, Eq.\,(\ref{4eq4}) implies that  spatial variation of the  Lorentz factor can produce electron density filaments when the beam is highly relativistic. The nonlinear terms left out in the linear analysis  can play an important role in generating the longitudinal fields in highly relativistic beams, when the amplitude of the FI becomes large.

\section{1D PIC Simulation Results}	
\label{s03}
\begin{figure}[h]
		\centering
		\vspace{-0pt}
			\includegraphics[width=7.5cm]{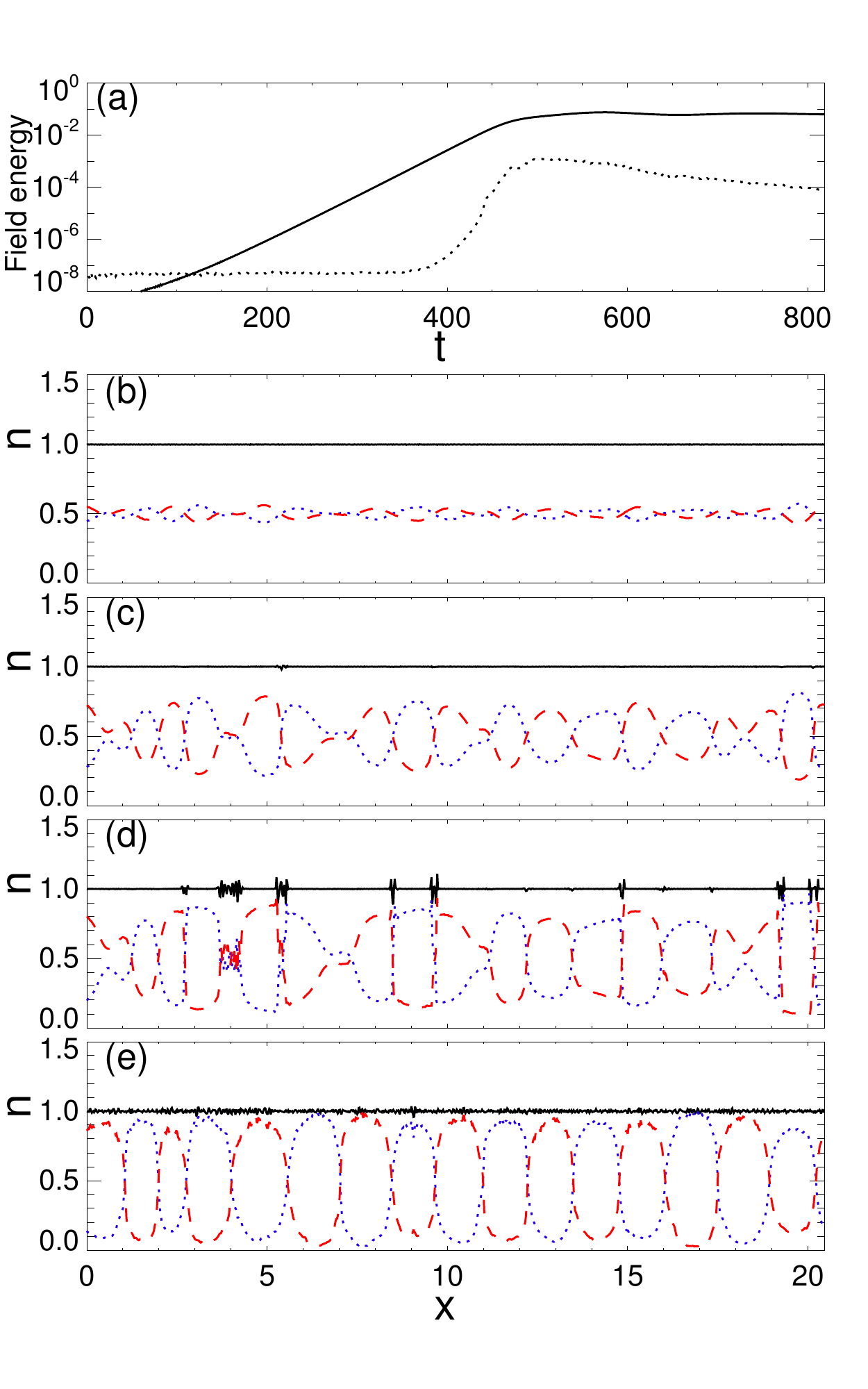}
		\vspace{-10pt}
		\caption{Nonrelativistic  counter streaming electron beams in 1D: No significant density filamention develops. (a) Field energy normalized by the total initial kinetic energy of electrons vs. simulation time: magnetic field energy $|B_y|^2$ (the solid line) and electric field energy $|E_x|^2$ (the dotted line); (b)\textendash{(e)} charge density distributions at different times, $e^-_1$ density (the dotted blue line), $e^-_2$ density (the dashed red line), and the total electron density (the solid black line): (b) at $t=348.04$, (c) at $t=435.05$, (d) at $t=460.64$, and (e) at $t=818.91$.}\label{nonall}
	\end{figure}

	To confirm Eq. (\ref{4eq4}), 1D PIC simulations were  performed using the EPOCH code with periodic boundary conditions. \citep{Huynh2014, Huynh2016}  The code used the SI units. In the simulations, the number of computational particles (CPs) is set as  $10^3$  CPs/cell  for the ion population, and  $9.6\times 10^4$ CPs/cell/species for each electron species, which guarantees a good fluid behavior of the particles. The ion background uniformly spans the whole space with the density $n_i=10^6$/$m^3$, which corresponds to  $n_i=n_0=1$  in the normalized unit. The rest of the simulation parameters are presented in the normalized units. The thermal velocity of the electron was  $u_{th}=2.5\times 10^{-3}$, grid space $\Delta x=2\times 10^{-2} \lambda_e$, where $\lambda_e$ is the skin depth. The simulation domain  was $L_x=1024\, \Delta x$. Simulations were performed in both the nonrelativistic and relativistic regimes for the symmetric counter streaming electron beams with $n_{01}=n_{02}=n_i/2=0.5$. 
	
	In the nonrelativistic regime, perfectly counter streaming flow velocities $u_{01}=-u_{02}=u_0=2.5\times 10^{-2}$  were used in the simulation with the time step $\Delta t=2.5\times 10^{-3}$.   The linear stage of the instability stops at approximately  $t\approx 435$ (Fig. \ref{nonall}a). During the linear stage, two electron beams   develop filamentation at the same rate. In Figs. \ref{nonall}b and \ref{nonall}c, when $e^-_1$ particles (the dotted blue line) are pinched, the same amount of $e^-_2$ particles (the dashed red line) are expelled.  For example, it can be seen in Fig. \ref{nonall}c that $e^-_1$ density (the dotted blue line) and  $e^-_2$ density (the dashed red line)  fluctuate around 0.5 opposite to each other. This confirms that $\partial (p_{x1}+p_{x2})/\partial x =0$ is satisfied.  Therefore, the total electron density  remains at $n=1$, which is indicated as the solid black lines in  Figs. \ref{nonall}b and  \ref{nonall}c.

    In the nonlinear stage ($t>435$),  tiny local perturbations of the total electron density are  seen as shown in Fig. \ref{nonall}d,  which are related to a sudden increase in the electric field energy $|E_x|^2$ in Fig. \ref{nonall}a.  The growth rate of $|E_x|^2$ is  2 times that of $|B_y|^2$. In a fully saturated stage,  Fig. \ref{nonall}e shows that magnitudes of  the total electron density perturbations are very small (the solid black line in Fig. \ref{nonall}e). Thus, on the whole, the total electron density remains constant in the nonrelativistic regime.

	\begin{figure}[h]
		\centering
		\vspace{-0pt}
			\includegraphics[width=9.5 cm]{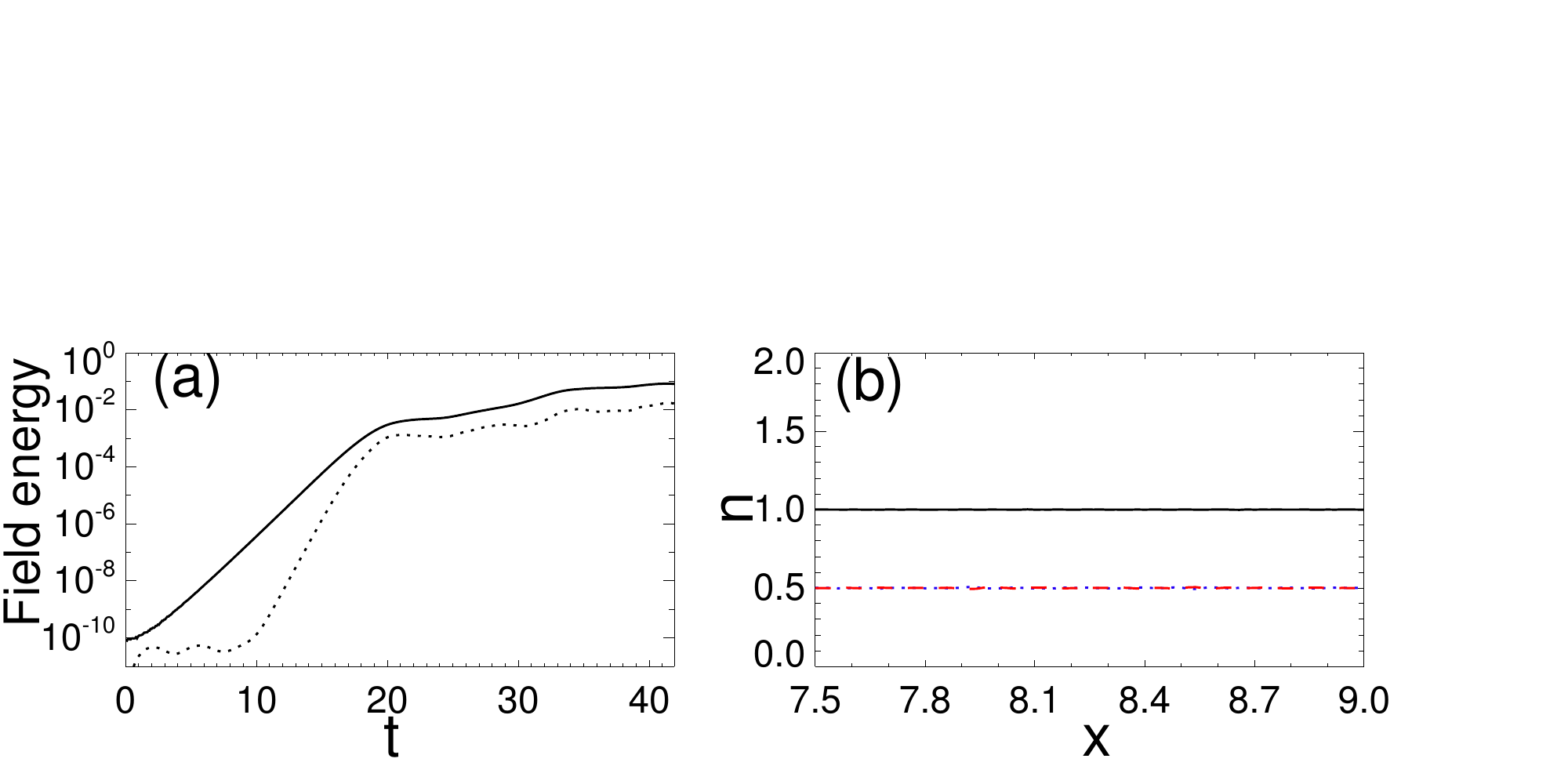}
		\caption{Simulation results  for the relativistic counter streaming electron beams. (a) Field energy normalized by the total initial kinetic energy of electrons vs. simulation time: magnetic field energy $|B_y|^2$ (the solid line) and electric field energy $|E_x|^2$ (the dotted line); (b) zoom-in charge density distribution at the beginning of linear stage ($t=6.4$): $e^-_1$ density (the dotted blue line), $e^-_2$ density (the dashed red line), and the total electron density (the solid black line). $e^-_1$ and $e^-_2$ lines exactly overlap at $n=0.5$. Density filamentation has not yet been developed.} \label{relin}
	\end{figure}

The simulations performed in the relativistic regime also used the perfectly counter streaming flow velocities $u_{01}=-u_{02}=u_0=0.6$ with the time step $\Delta t = 10^{-3}$.   	Contrary to the nonrelativistic case, the simulation results in the relativistic regime  evidence that the electron density filamentation can  occur. The linear stage of the instability ends around $t \approx 18$ (Fig. \ref{relin}a).   At the beginning of the linear  phase ($t<10$), there is no change in total electron density (the solid black line in Fig. \ref{relin}b). Because   the velocity distribution is uniform at this time,   filamentation does not occur.

\begin{figure}[t]
		\vspace{-0pt}
		\centering
			\includegraphics[width = 10cm]{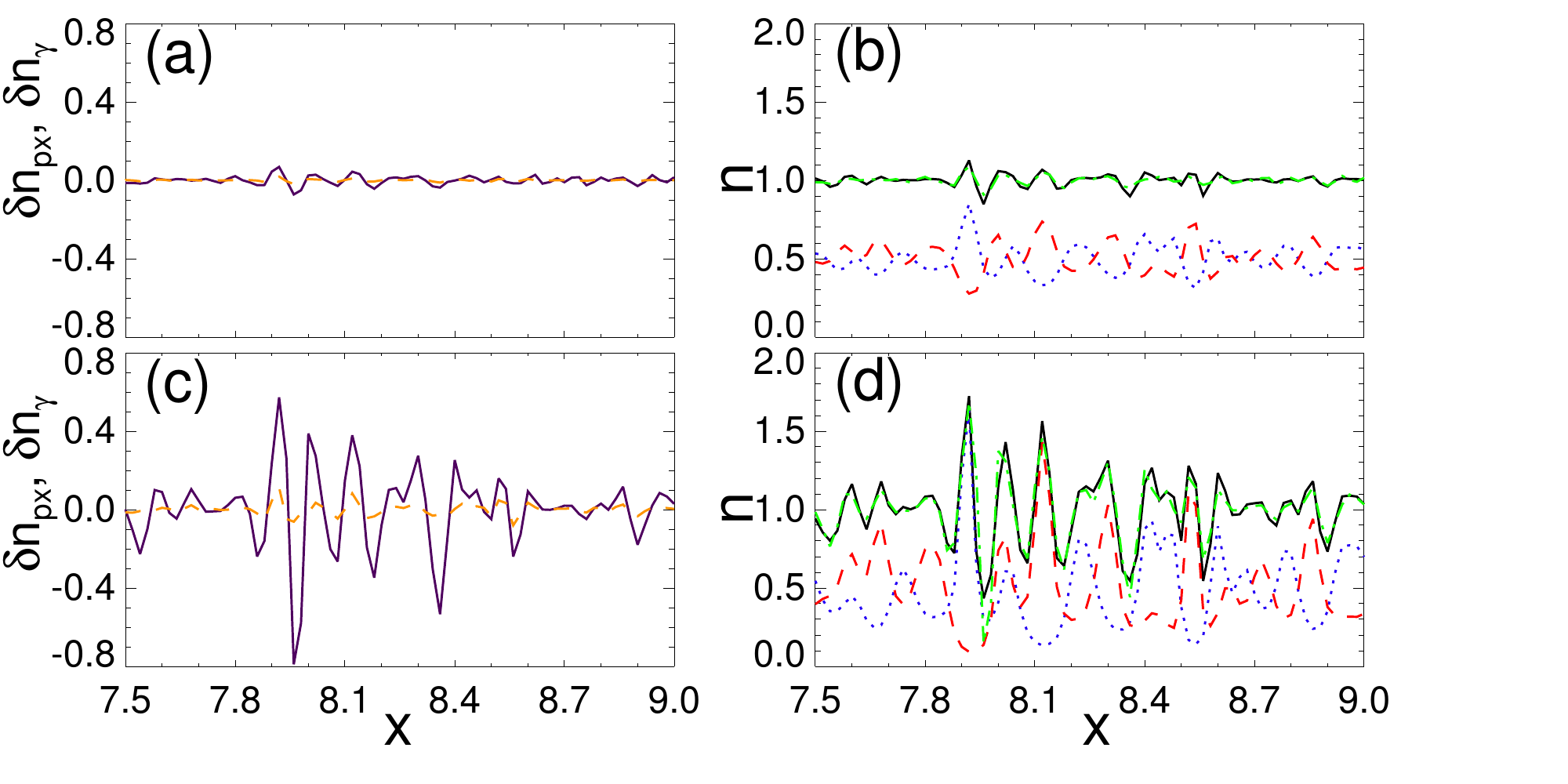}
		\caption{Zoom in of the  simulation  results  for the  relativistic  counter streaming electron beams in the time interval between  the middle ($t=14.87$) and near the end ($t=17.1$) of the linear stage, where filamentation develops. (a) at $t=14.87$ and (c) at $t=17.1$ are  theoretical estimations of two terms in Eq.\,(\ref{4eq4}) with fluid quantities $p_{1x},\, p_{2x},\, \gamma_1,$ and $\gamma_2$ measured from the simulation: $\delta n_{px}$ (the dashed orange line) and $\delta n_{\gamma}$ (the solid purple line);  (b) at $t=14.87$ and (d) at $t=17.1$ are the electron density in $x$ obtained from the simulation: $e^-_1$ density (the dotted blue line), $e^-_2$ density (the dashed red line), and the total electron density (the solid black line). Theoretical estimations of the total electron density $n=1+\delta n_{px}+\delta n_{\gamma}$ are over plotted using the dashed green lines in (b) and (d). The dashed green and solid black lines almost overlap.  Comparison between the theoretical analysis and simulation results shows good agreement.} \label{relno}
	\end{figure}

From the middle to the end of the  linear stages,   the electron density filamentation starts to build up. If we define $\delta n_{px}\equiv \Gamma_{max}\partial (p_{x1}+p_{x2})/2\partial x$ and $\delta n_{\gamma} \equiv \partial^2 (\gamma_{1}+\gamma_{2})/2\partial x^2$,  Eq.\,(\ref{4eq4})  can be simply rewritten as $n=1+ \delta n_{px} +\delta n_{\gamma}$.  		To see the  contributions from $\delta n_{px}$ and $\delta n_{\gamma}$, we measured the fluid quantities $p_{x1},\, p_{x2}, \, \gamma_{1}, \, \gamma_{2}$ from the simulation, and then put these values in Eq.\,(\ref{4eq4}). Figures \ref{relno}a and \ref{relno}c,  at $t=14.87$ and $t=17.1$, show that the amplitudes of $\delta n_{\gamma}$ (the solid purple lines in Figs. \ref{relno}a and \ref{relno}c) are much higher than those of $\delta n_{px}$ (the dashed orange lines in Figs. \ref{relno}a and \ref{relno}c)  given  $\partial^2 (\gamma_{1}+\gamma_{2})/\partial x^2 \neq 0$,   according  to Eq.\,(\ref{4eq4}). Therefore,  the electron density filamentation occurs mainly by the spatial variation of the Lorentz factor.  As can be seen in Fig. \ref{relno}b,  many sharp peaks in the total electron density appear (the solid black line in Fig. \ref{relno}b). Equation (\ref{4eq4}) is plotted using the dashed green line in Fig. \ref{relno}b. There is a good agreement between the simulation result (the solid black line) and the theoretical prediction (the dashed green line). The two lines  almost overlap.
Due to the rapid change of $\delta n_{\gamma}$, the magnitudes of  peaks in Fig. \ref{relno}b quickly increase within a few plasma frequencies. For example, at $x \approx 7.9$, the maximum value of $n\approx 1.15$ at $t=14.87$ in Fig. \ref{relno}b  increases to $n \approx 1.75$ at $t = 17.1$ in Fig. \ref{relno}d. At this time ($t=17.1$, near the end of the linear stage), the analytical estimation (the dashed green line in Fig. \ref{relno}d) agrees with the simulation results (the solid black line in Fig. \ref{relno}d).

			\begin{figure}[h]
		\vspace{-0pt}
		\centering
			\includegraphics[width=7.5 cm]{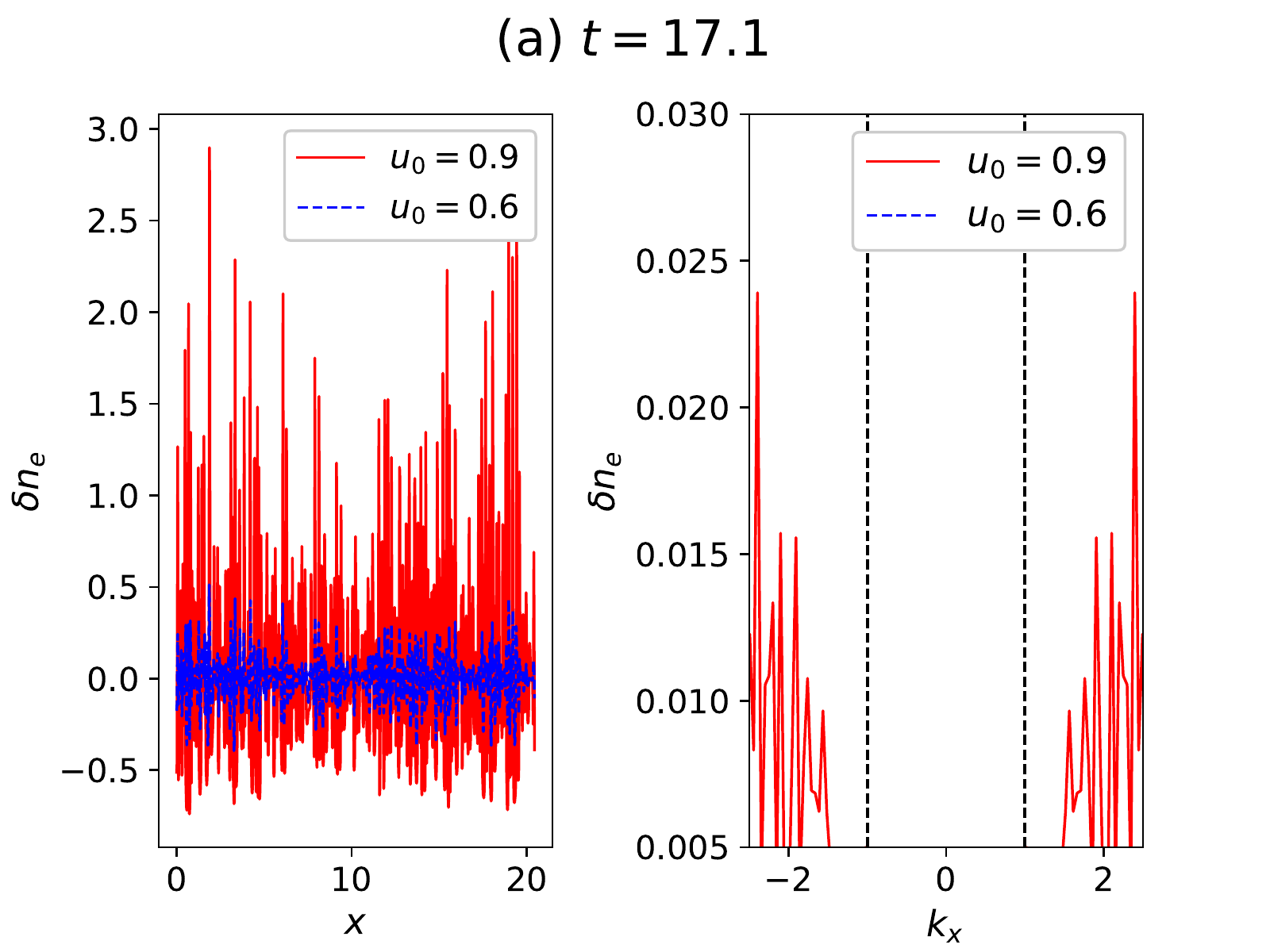}
			\includegraphics[width=7.5 cm]{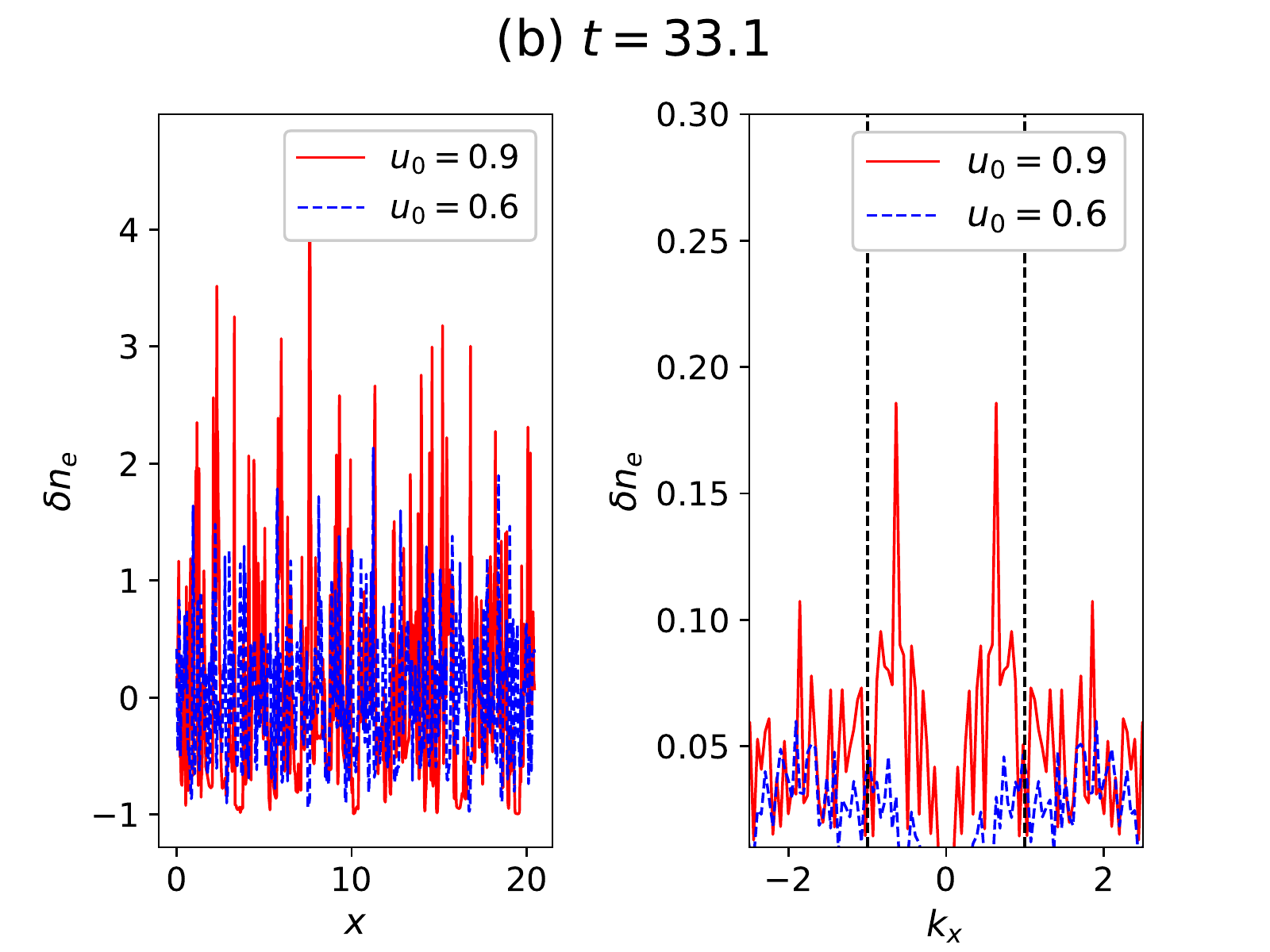}
		\caption{1D density filamentation in real space and FFT spectral domains for   velocities  $u_0=0.6$  and  $0.9$ at two differnet times, (a) $t=17.1$ and (b)  $t=33.1$ . Peak intensities in real and Fourier space increase as the velocity increases.  The dotted lines at $k_x=\pm 1$ in the FFT spectra represent the Langmuir modes.  The spectral behavior of density filaments  indicates the evolution into larger spatial scales $k_x <1$, into long-wavelength modes  than the Langmuir mode.  } \label{1dfft}
	\end{figure}

	\begin{figure}[h]
		\vspace{-0pt}
		\centering
			\includegraphics[width=7.5 cm]{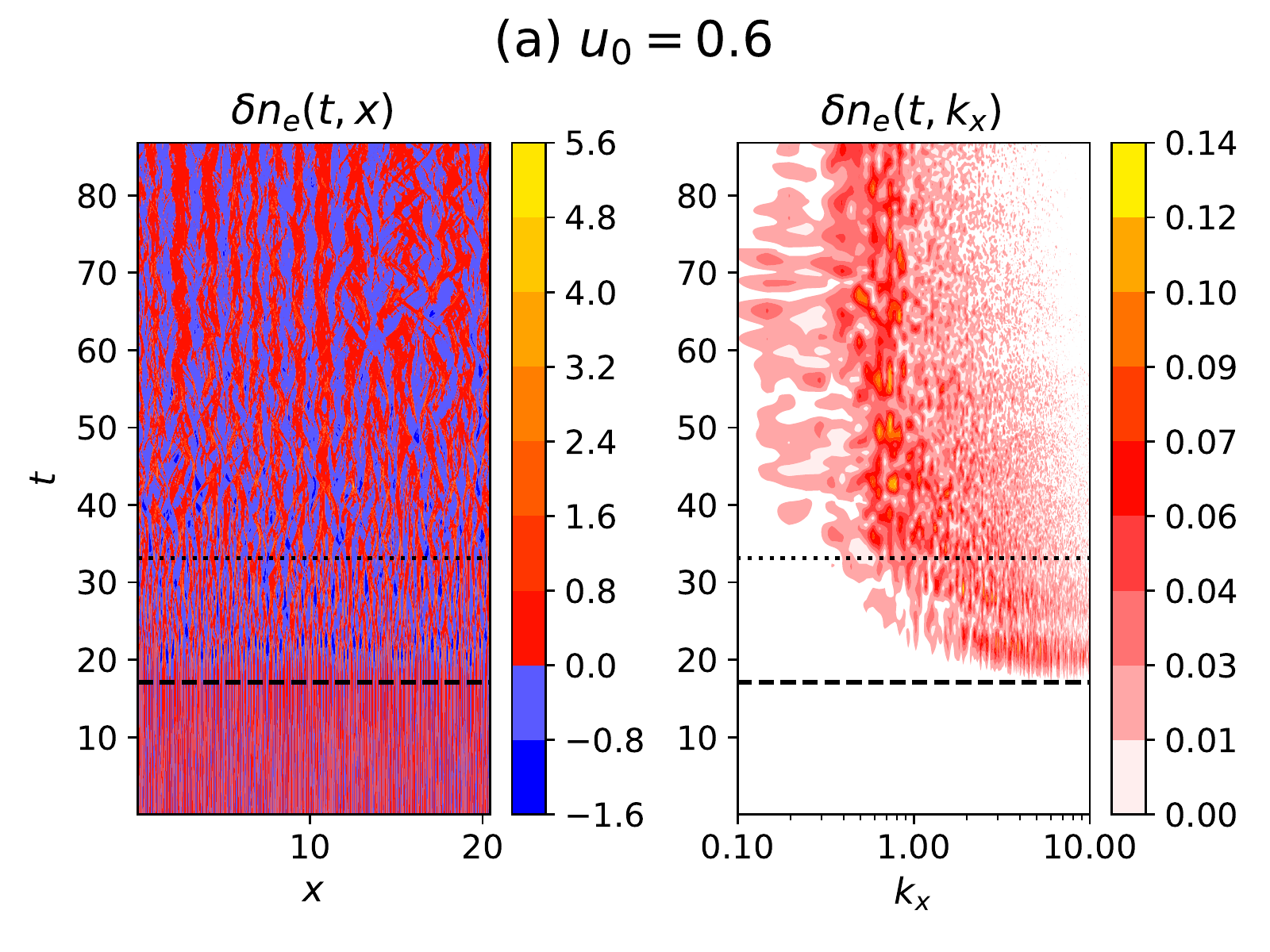}
			\includegraphics[width=7.5 cm]{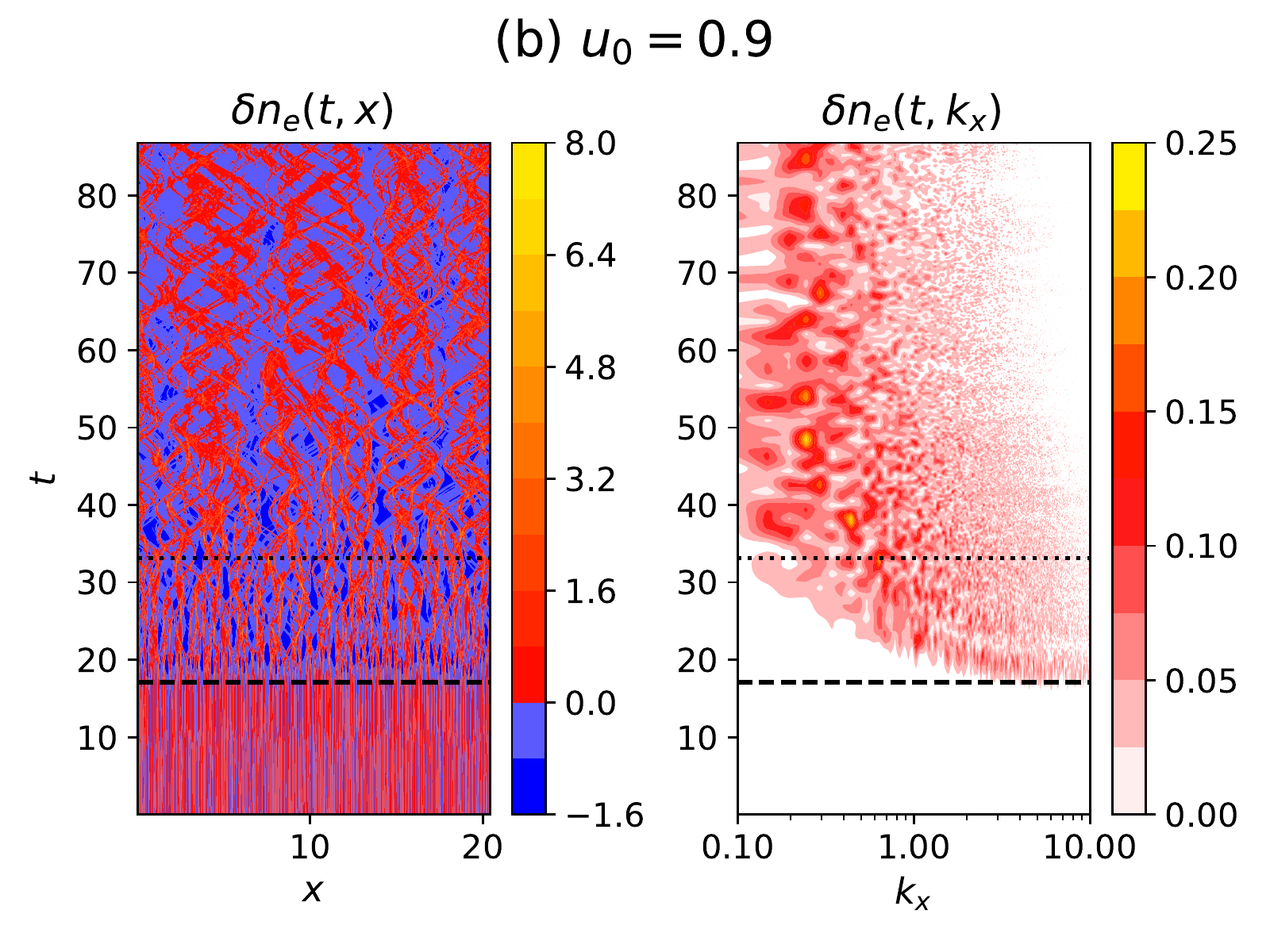}
		\caption{Time evolution of density filaments and  their Fourier spectra for (a) $u_0=0.6$  and (b) $u_0=0.9$.  Merging to longer filaments can be seen. Dashed and dotted horizontal lines mark the times at $t=17.1$ and $t=33.1$ used in Fig.\,\ref{1dfft}, respectively .} \label{1d_evol}
	\end{figure}

 Near the end of the linear stage $t \approx 18$ (Fig. \ref{relin}a),  the electrostatic field becomes stronger due to the space-charge separation increase, so that the electrostatic field energy becomes a fraction of the magnetic field energy.
Figures \ref{1dfft}a and \ref{1dfft}b illustrate the development of large-scale   density filaments at two different times, (a) $t=17.1$ and (b) $t=33.1$. For a higher relativistic velocity $u_0=0.9$, a higher peak density and a smaller wave number corresponding to a greater spatial scale can be seen than in the case for $u_0=0.6$.  Figure \ref{1d_evol} shows the time evolution of density filaments in the real space and in the spectral domain for $u_0=0.6$ and $u_0=0.9$. In the early linear phase,  many short wavelength modes appear, but in time, filamentation merging takes place and contribution from short wavelength modes diminishes.  A tendency  to merge into a longer wavelength filament mode than the Langmuir mode is seen in the spectral domain for the wavenumber $k_x$ less than $1$.

\begin{figure}[h]
		\centering
			\includegraphics[width=9cm]{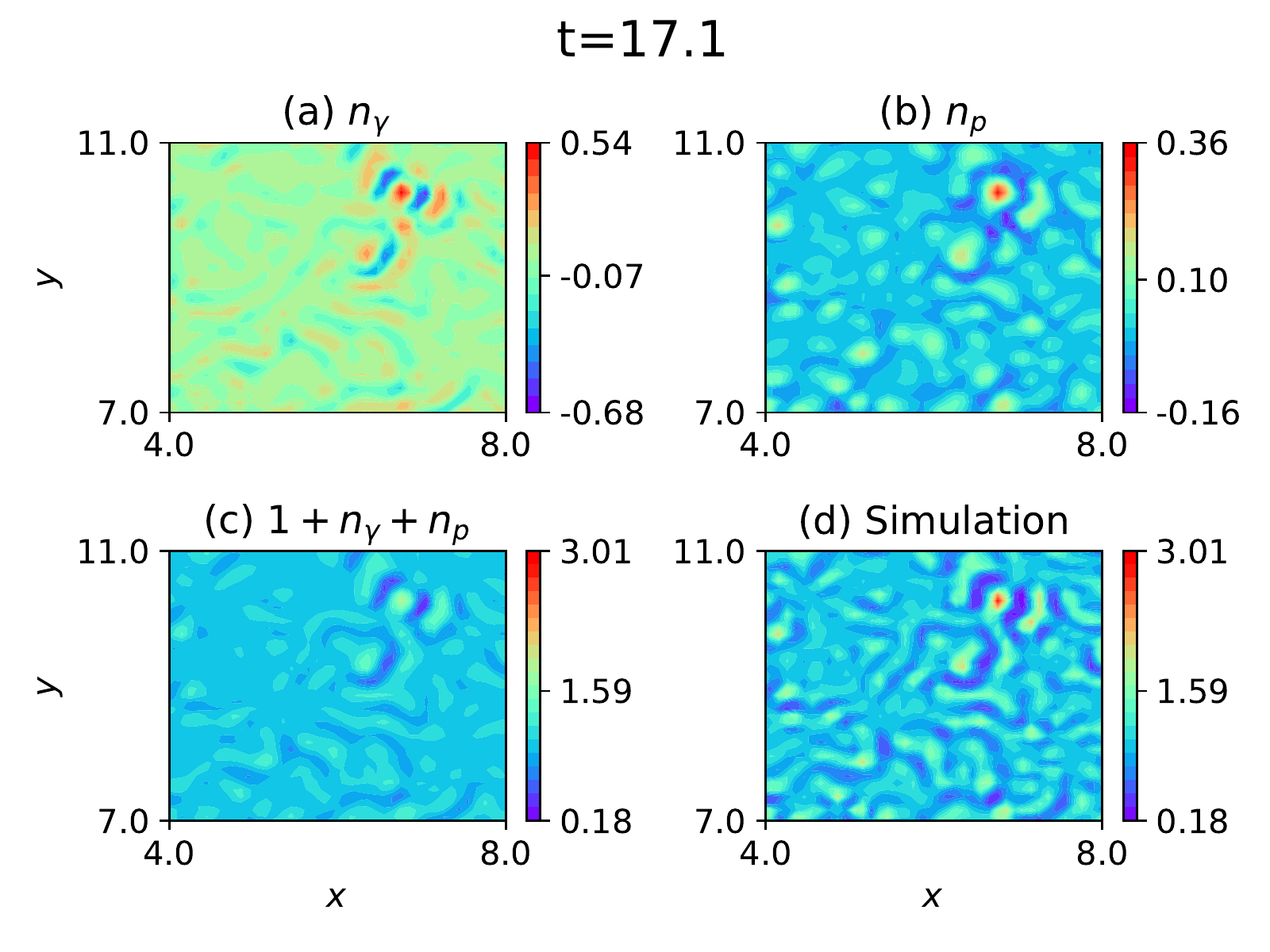}
		\caption{2D zoom-in simulation results, showing filamentation for  relativistic counter streaming  electron beams with $u_0=0.6$ at $t=17.1$: (a) $ n_{\gamma}$, (b) $ n_{p}$, (c) the total electron density are estimated using Eq.\,(\ref{5eq11}),  and (d) is the electron density from simulation. The patterns  of (c)   and (d)  are similar.}\label{2drelno}
	\end{figure}

\section{Density Filamentation by the Lorenz factor in 2D simulations}
\label{s04}
In the 2D simulation, similar parameters to the one for the 1D study are adopted. The number of computational particles (CPs) is taken as   $2.5 \times 10^2$  CPs/cell  for the ion population, and  $2.4\times 10^4$ CPs/cell/species for each electron species. As before, the ion background uniformly spans the whole space with  density $n_i=10^6$/$m^3$, which corresponds to  $n_i=n_0=1$  in the normalized units.  The thermal velocity of the electron is $u_{th}=2.5\times 10^{-3}$, grid space $\Delta x= \Delta y = 0.1 \lambda_e$, and the simulation domain  $L_x = L_y =128\, \Delta x$.  In simulations, the symmetric counter streaming electron beams with $n_{01}=n_{02}=n_i/2=0.5$ are used, and periodic boundary conditions are taken in the $x$ and $y$ directions.

In the two-dimensional (2D)  geometry, it is difficult to separate the linear growth of filaments from the nonlinear merging process because they usually occur simultaneously. \citep{Rowlands2007,Dickmann2009}  In the nonlinear filament merging processes, the term $\partial (p_{x1}+p_{x2})/2\partial x$ can induce density filaments.  In 2D, Eq.\,(\ref{4eq4}) becomes

\begin{eqnarray}
		n &=&  1+\Gamma_{max}\frac{\partial (p_{x1}+p_{x2})}{2 \partial x} + \Gamma_{max}\frac{\partial (p_{y1}+p_{y2})}{2 \partial y} \nonumber \\
          &+& \frac{\partial^2 (\gamma_{1}+\gamma_{2})}{2 \partial x^2} + \frac{\partial^2 (\gamma_{1}+\gamma_{2})}{2 \partial y^2} \nonumber \\
          &=& 1 + n_{p} + n_{\gamma}, \label{5eq11}
	\end{eqnarray}
where $n_p \equiv \Gamma_{max}(\partial(p_{x1}+p_{x2})/2\partial x  + \partial(p_{y1}+p_{y2})/2\partial y) $ and $n_{\gamma} \equiv \partial^2 (\gamma_{1}+\gamma_{2}) / 2\partial^2 x + \partial^2 (\gamma_{1}+\gamma_{2}) / 2\partial^2 y$.

The 2D simulation results are shown in Fig. \ref{2drelno}. 2D zoom-in  results are plotted,  showing the analysis for the filamentation by weakly relativistic counter streaming  electron beams with the velocity $u_0=0.6$ at $t=17.1$. To plot the figures, $p$ and the Lorentz factor $\gamma$ are obtained at each grid points from the PIC simulation, and  the derivatives of these values  $n_p$ and $n_\gamma$ are taken using Eq. (19). The obtained $n_p$ and $n_\gamma$ and the sums of $1+n_\gamma +n_p$ are plotted on panels (a), (b) and (c), respectively. Then, the density measured from the simulation is plotted on panel (d). Overall patterns of panels (c) and (d) match well with each other.    The figures says that the perturbation of the total electron density is mainly governed by $n_{\gamma}$ (Fig. \ref{2drelno}a).   This result seems to imply that the essential physics discussed  in  1D has not changed in 2D.

			\begin{figure}[h]
		\vspace{-0pt}
		\centering
			\includegraphics[width=9cm]{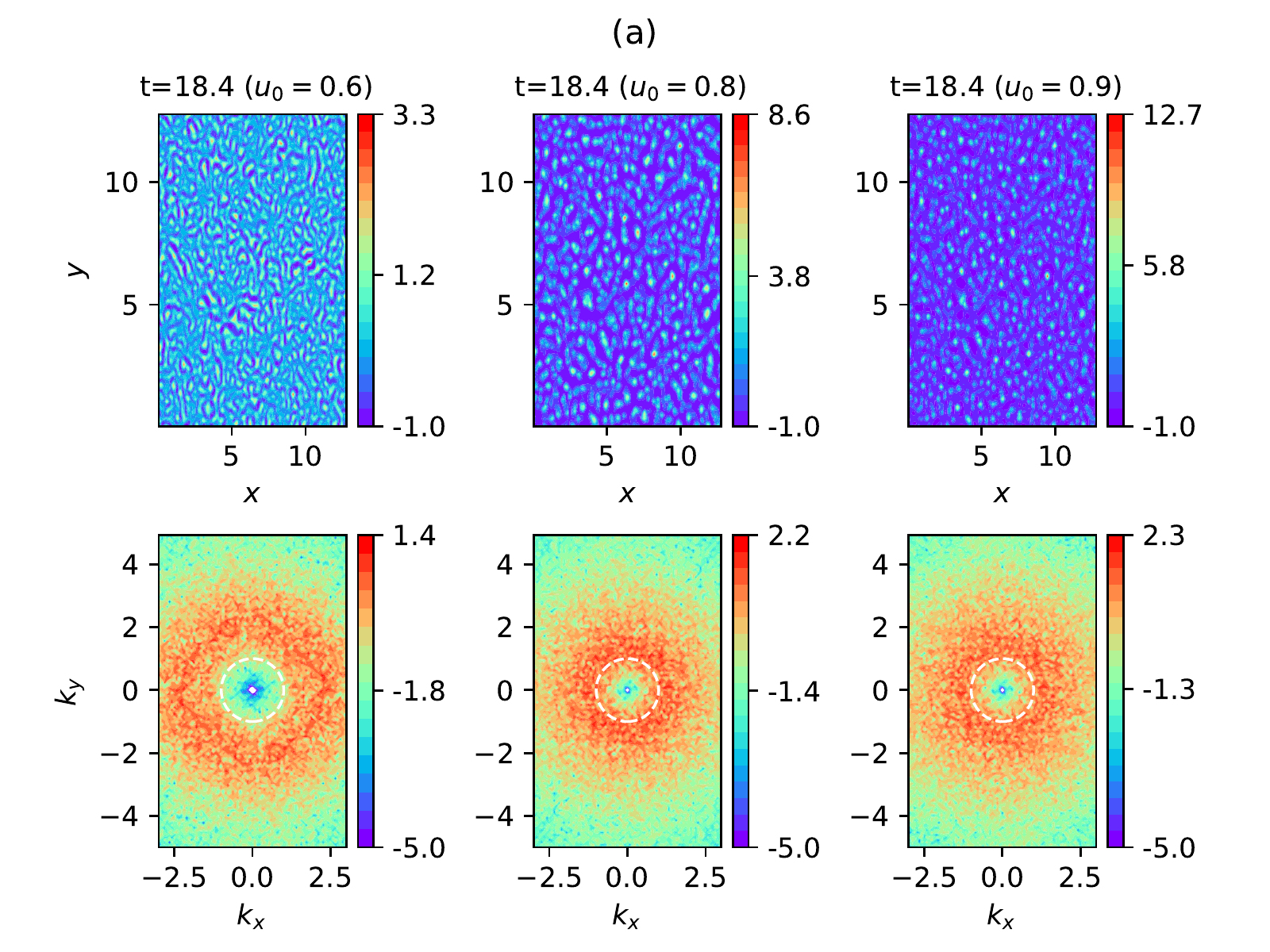}
			\vspace{2pt}
			\includegraphics[width=9cm]{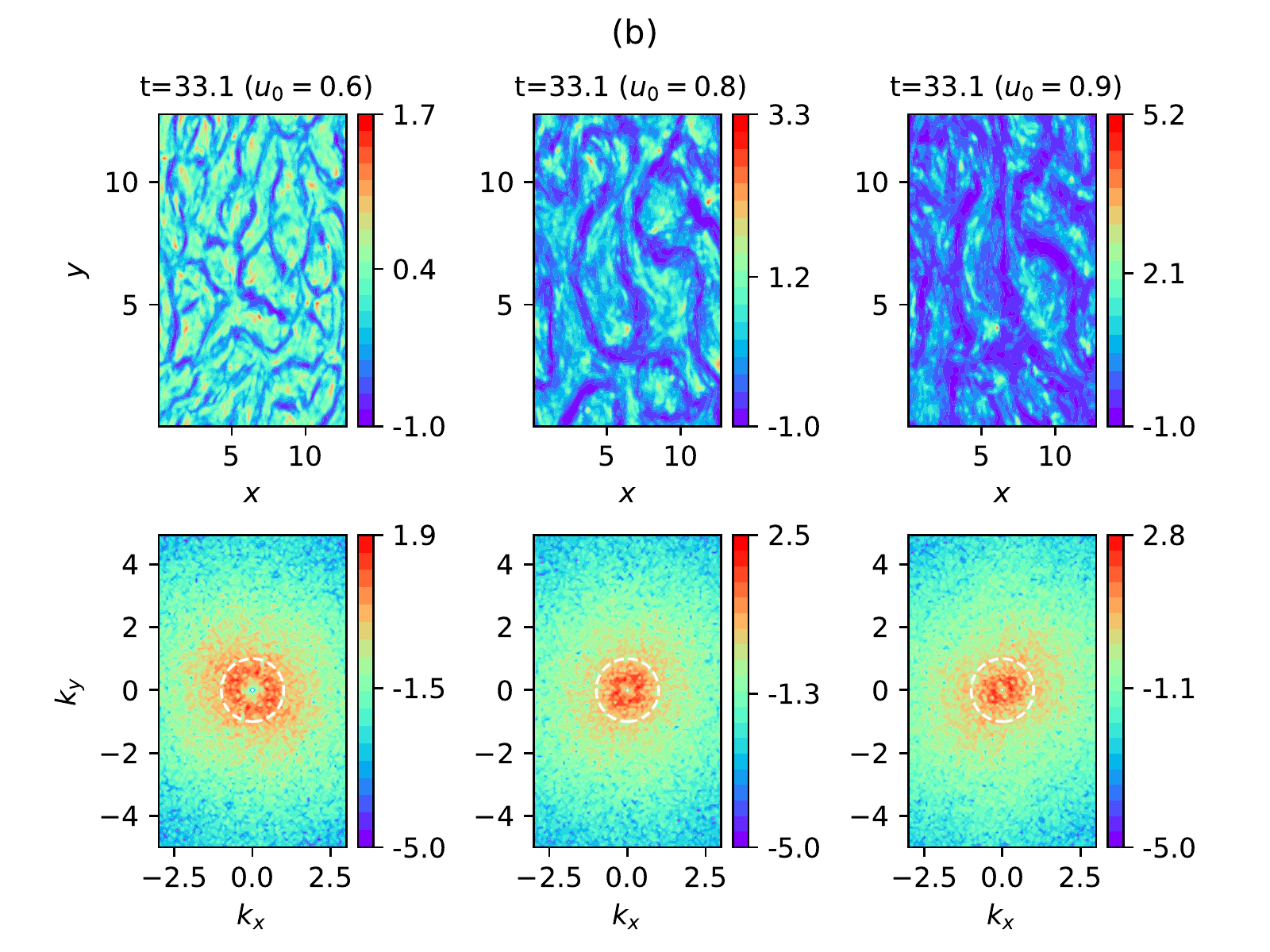}
		\caption{Density filamentation in the real space and FFT spectral domains for  different relativistic velocities  $u_0=0.6$, $0.8$ and $0.9$ at two different times, (a) $t=18.4$ and (b) $t=33.1$. The scale bar denoting the peak intensity in real space drastically increases with the velocity increase.  The white circles in the FFT spectra represent the Langmuir modes.   } \label{2dfft}
	\end{figure}


	In Figure \ref{2dfft},   density filaments and  FFT spectra of 2D simulation results are shown for three different relativistic velocities of $u_0=0.6, 0.8$ and $0.9$  at two different times, $t=18.4$ and $t=33.1$. As in the one-dimensional analysis, for a higher relativistic velocity, the maximum density increases to a higher value, and the  spatial scale increases.  The circles in the FFT spectra denote the Langmuir modes.  For  $u_0=0.9$, the color index bar for the density indicates that the maximum value of the density filament becomes 4 to 5 times greater than that in the case of $u_0=0.6$, while  density filaments  merge into the ones with a larger length scale  than the Langmuir modes, filling the circle quickly.

\section{Conclusion}
\label{s05}

Density filamentation  has been observed in many astrophysical and laser-driven plasma simulations, where transversal electromagnetic fields are prevalent.  To explain it, several mechanisms for the generation of the electrostatic field have been proposed. Among these, obliquely propagating mixed mode is currently the dominant idea for such electrostatic field generation.   {
In this paper, we have examined, taking a slightly different approach, the  generation of the electrostatic field via a nonlinear process of the FI, generalizing Rowlands et al.'s observation to a relativistic situation.  We were able to demonstrate that the nonlinear effects of the  Weibel/filamentation instability in a relativistic  beam-plasma system can indeed give rise to density filamentation and electrostatic field generation.     We presented  a theory together with the supporting  1D and 2D PIC simulation results.
 
The Lorentz factor distribution essentially describes how the fluid velocity varies in space and can be directly related to the nonlinear fluid motion.  According to the conventional linear analysis,  only the transverse fields are  possible in perfectly counter streaming beam plasmas, whether the beams are relativistic or nonrelativistic. 
The nonlinear fluid motion,  which can be compactly expressed as the gradient of the Lorentz factor, is shown to bring in a new feature of the electrostatic field and density filamentation not possible in the linear analysis. 
 The nonlinearly driven density filamentation depends on the magnitude of the relativistic $\gamma$ and its nonuniformity, which seems to have important implications to highly relativistic plasmas. The nonlinear behavior of the filamentation instability can be related to the particle velocity distribution in space via the gradient of the Lorentz factor,  inducing the electrostatic field.

In many astrophysical and laser-driven plasmas, density filamentation and the associated electrostatic fields are important for particle heating and acceleration.   It is still necessary to identify the particle heating and acceleration mechanisms.  The results in this paper  could be useful  in understanding the physics of relativistic astrophysical and laboratory plasmas more deeply,  for instance,  the synchrotron radiation in  gamma ray bursts or the particle heating and  acceleration in collisionless shocks.


\begin{acknowledgements}
This work was supported by the grant from the Institute for Basic Science, Korea (IBS-R012-D1).
\end{acknowledgements}

\appendix
\section{Derivation of relativistic momentum equation with the Lorentz factor}
\label{Appendix}
From the relativistic momentum-energy relationship 
\begin{eqnarray}
 E^2 = p^2 c^2 + (mc^2)^2  
 \label{eq_A1}
 \end{eqnarray}
 with $E = \gamma mc^2$,  ${\bf p} = \gamma m{\bf v}$,  and $\gamma_\alpha^2 = 1+{\bf p}_\alpha^2$ in the normalized units. \\

The derivative of this equation gives 
\begin{eqnarray}
\nabla \gamma_\alpha^2 &=& \nabla (1+ {\bf p}_\alpha^2) = \nabla ({\bf p}_\alpha \cdot {\bf p}_\alpha ) \nonumber \\[12pt] 
&=&  2 (\gamma_\alpha {\bf u}_\alpha \times (\nabla \times {\bf p}_\alpha)) + 2 \gamma_\alpha ({\bf u}_\alpha \cdot \nabla) {\bf p}_\alpha,
\label{eq_A2}
\end{eqnarray}
where
$\nabla ({\bf p}_\alpha \cdot {\bf p}_\alpha)$ is replaced with
\begin{eqnarray}
\nabla ({\bf p}_\alpha \cdot {\bf p}_\alpha) &=& 2({\bf p}_\alpha \times (\nabla \times {\bf p}_\alpha)) + 2 ({\bf p}_\alpha \cdot \nabla) {\bf p}_\alpha \nonumber \\[12pt] 
&=& 2 (\gamma_\alpha {\bf u}_\alpha \times (\nabla \times {\bf p}_\alpha)) + 2 \gamma_\alpha ({\bf u}_\alpha \cdot \nabla) {\bf p}_\alpha,
\label{eq_A3}
\end{eqnarray}
by using the vector identity,
\begin{eqnarray}
\nabla ({\bf A} \cdot {\bf B}) &=& {\bf A} \times (\nabla \times {\bf B}) + {\bf B} \times (\nabla \times {\bf A}) \nonumber \\[12pt] 
&+& ({\bf A} \cdot \nabla) {\bf B} + ({\bf B} \cdot \nabla ) {\bf A}.
\label{eq_A4}
\end{eqnarray}
As a result, we obtain
\begin{eqnarray}
\nabla \gamma_\alpha = {\bf u}_\alpha \times (\nabla \times {\bf p}_\alpha) + ({\bf u}_\alpha \cdot \nabla) {\bf p}_\alpha.
\label{eq_A5}
\end{eqnarray}
Taking the curl of Eq.\,(\ref{4eq1}), we have
\begin{eqnarray}
\frac{\partial (\nabla \times \textbf{p}_{\alpha})}{\partial t} &+& \nabla \times \left[ \nabla \gamma_\alpha - {\bf u}_\alpha \times (\nabla \times {\bf p}_\alpha) \right] \nonumber \\[12pt] 
&=& - \frac{\partial (\nabla \times {\bf E})}{\partial t} - \nabla \times [{\bf u}_\alpha \times {\bf B}],
\label{eq_A6}
\end{eqnarray}
which becomes
\begin{eqnarray}
\frac{\partial (\nabla \times \textbf{p}_{\alpha}) }{\partial t} &-& \nabla \times \left[ {\bf u}_\alpha \times (\nabla \times {\bf p}_\alpha ) \right] \nonumber \\[12pt] 
&=& \frac{\partial {\bf B}}{\partial t} - \nabla \times [{\bf u}_\alpha \times {\bf B}],
\label{eq_A7}
\end{eqnarray}
using Faraday's law $\nabla \times {\bf E} = - \partial {\bf B} / \partial t$.
When the R.H.S. terms are moved to the left, 

\begin{eqnarray}
\frac{\partial ( (\nabla \times \textbf{p}_{\alpha})  - {\bf B}) }{\partial t} &-& \nabla \times \left[ {\bf u}_\alpha \times \left\{(\nabla \times  {\bf p}_\alpha ) - {\bf B} \right\} \right] = 0.
\label{eq_A7_add}
\end{eqnarray}

\begin{equation}
\frac{\partial \mathbf{\Omega}_\alpha}{\partial t} -\nabla \times (\mathbf{u}_\alpha \times \mathbf{\Omega}_\alpha)=0,
\end{equation}
where $\mathbf{\Omega}_\alpha = \mathbf{B}-\nabla \times \mathbf{p}_\alpha$. Initially, the velocities of counterstreaming flows are uniform and thus $\nabla \times \mathbf{p}_\alpha =0$, and there is no magnetic field. As a result, $\mathbf{\Omega}_\alpha = 0$ initially. Then,    $\mathbf{\Omega}_\alpha$ remains zero through out the time  (See Appendix B of Ref. [\onlinecite {Kaganovich2001}] ).  This leads to the London equation $\mathbf{B}=\nabla \times \mathbf{p}_\alpha$.

 Putting back ${\bf B} = \nabla \times {\bf p}_\alpha$ into Eq.\,(\ref{4eq1}), we have 
\begin{eqnarray}
\frac{\partial \textbf{p}_{\alpha}}{\partial t} &+& \left(\textbf{u}_{\alpha} \cdot \nabla \right) \textbf{p}_{\alpha} \nonumber \\[12pt] 
&=& -\textbf{E} - \textbf{u}_{\alpha}\times\textbf{B} = -\textbf{E} - \textbf{u}_{\alpha} \times (\nabla \times {\bf p}_\alpha),
\label{eq_A8}
\end{eqnarray}
which can be reduced to 
\begin{eqnarray}
\frac{\partial \textbf{p}_{\alpha}}{\partial t} + \left(\textbf{u}_{\alpha} \cdot \nabla \right) \textbf{p}_{\alpha} + \textbf{u}_{\alpha} \times (\nabla \times {\bf p}_\alpha)= -\textbf{E}. 
\label{eq_A9}
\end{eqnarray}
This equation  can be finally written  with the help of Eq.\,(\ref{eq_A5}) as
\begin{eqnarray}
\frac{\partial \textbf{p}_{\alpha}}{\partial t} &=& - [\left(\textbf{u}_{\alpha} \cdot \nabla \right) \textbf{p}_{\alpha} + \textbf{u}_{\alpha} \times (\nabla \times {\bf p}_\alpha)] -\textbf{E} \nonumber \\[12pt] 
&=&- \nabla \gamma_\alpha -\textbf{E}. 
\label{eq_A10}
\end{eqnarray}
By using London's equation, Eq.\,(\ref{eq_A5}) can also be written as 
\begin{eqnarray}
\nabla \gamma_\alpha =  {\bf u}_\alpha \times {\bf B} + ({\bf u}_\alpha \cdot \nabla) {\bf p}_\alpha.
\label{eq_A11}
\end{eqnarray}

\bibliographystyle{aipnum4-2.bst}
\bibliography{RFI.bib}

\end{document}